\documentclass[a4paper,11pt]{article}
\usepackage{pos}

\usepackage{mathtools}

\newcommand{\D}{\text{d}}

\newcommand{\q}{\left(}
\newcommand{\ie}{\emph{i.e.}}

\newcommand{\w}{\right)}

\newcommand{\TFI}{T_{\rm FI}}

\newcommand{\TSW}{T_{\rm SW}}

\newcommand{\qnav}{\langle q^n\rangle}

\newcommand{\gs}{g_{*S}}

\newcommand{\class}{\textsc{class}}

\title{Cosmological imprints of non-thermalized dark matter}

\author[a]{Quentin Decant}
\emailAdd{quentin.decant@ulb.ac.be}
\author*[b,c]{Jan Heisig}
\emailAdd{heisig@physik.rwth-aachen.de}
\author[d]{Deanna C. Hooper}
\emailAdd{deanna.hooper@helsinki.fi}
\author[a,e]{Laura Lopez-Honorez}
\emailAdd{llopezho@ulb.ac.be}

\affiliation[a]{Service de Physique Th\'eorique, CP225, Universit\'e Libre de Bruxelles, Bld du Triomphe, B-1050 Brussels, Belgium}
\affiliation[b]{Institute for Theoretical Particle Physics and Cosmology, RWTH Aachen University, Sommerfeldstr. 16, D-52056 Aachen, Germany}
\affiliation[c]{Centre for Cosmology, Particle Physics and Phenomenology (CP3), Universit\'e catholique de Louvain, B-1348 Louvain-la-Neuve, Belgium}
\affiliation[d]{Department of Physics and Helsinki Institute of Physics, PL 64, FI-00014 University of Helsinki, Finland}
\affiliation[e]{Theoretische Natuurkunde, Vrije Universiteit Brussel and The International Solvay Institutes, Pleinlaan 2, B-1050 Brussels, Belgium}

\abstract{
Non-thermalized dark matter is a cosmologically viable alternative to the widely studied  weakly interacting massive particle. We study the evolution of the dark matter phase-space distributions arising from freeze-in and superWIMP production as well as the combination of both. Utilizing our implementation in \class, we investigate the cosmological imprints on the matter power spectrum, constrained by Lyman-$\alpha$ forest observations. For the explicit example of a colored $t$-channel mediator model, we explore the cosmologically allowed parameter space highlighting the interplay of Lyman-$\alpha$ constraints with those from Big Bang Nucleosynthesis and the LHC\@. 
}

\FullConference{%
  *** The European Physical Society Conference on High Energy Physics (EPS-HEP2021), ***\\
  *** 26-30 July 2021 ***\\
  *** Online conference, jointly organized by Universit{\"a}t Hamburg and the research center DESY ***
}

\begin{document}
\maketitle

\section{Introduction}
Cosmological observations imply that the formation of structures in our Universe is dominantly driven by dark matter (DM) making up 80\% of the total matter
content in our Universe~\cite{Planck:2018vyg}. Nevertheless, the microscopic properties of DM are still illusive.
Despite substantial experimental efforts, no clear hint for any non-gravitational interaction of DM with the standard model has been found, thereby imposing strong constraints on the respective interaction strength.

A complementary path to constraining particle DM models is the study of the clustering properties of matter. 
When DM is produced with a momentum distribution significantly different to the one of
cold DM, small-scale structures can be washed out by DM free-streaming
causing a cut-off in the matter power spectrum.
This effect can be probed by Lyman-$\alpha$ forest observations and has been interpreted
in the canonical warm DM scenario~\cite{Viel:2013fqw,Irsic:2017ixq,Palanque-Delabrouille:2019iyz},
excluding masses below 5.3 keV~\cite{Palanque-Delabrouille:2019iyz} (or 1.9 keV under considerably more conservative assumptions~\cite{Garzilli:2019qki}).

In this article, we summarize our recent results~\cite{Decant:2021mhj} re-interpreting Lyman-$\alpha$ constraints in models of non-thermalised DM, \ie~in a scenario where DM couples so weakly to the standard model that it never reaches thermal equilibrium with the primordial plasma. We consider freeze-in production from renormalizable operators, namely decays and scatterings of bath particles~\cite{McDonald:2001vt,Asaka:2005cn,Hall:2009bx}, as well as late decays of frozen-out mother particles, \ie~the superWIMP production~\cite{Covi:1999ty,Feng:2003uy}. 
For a dark sector containing DM and a (heavier) mediator particle, odd under a discrete $Z_2$-symmetry, both production mechanisms are present and can contribute to a similar amount. 

In Sec.~\ref{sec:Lyabounds}, we discuss Lyman-$\alpha$ constraints for a pure freeze-in and superWIMP scenario 
refining earlier results obtained the literature~\cite{Heeck:2017xbu,Boulebnane:2017fxw,Bae:2017dpt,Ballesteros:2020adh,DEramo:2020gpr}. In Sec.~\ref{sec:appl}, we consider the mixed freeze-in/superWIMP scenario and apply our analysis to a simplified DM model with a colored $t$-channel mediator improving on earlier results~\cite{Garny:2018ali}. We conclude our discussion in Sec.~\ref{sec:concl}.

\section{Dark matter momentum distribution and Lyman-$\alpha$ constraints}\label{sec:Lyabounds}

The impact of DM on structure formation depends on its momentum distribution $f_\chi(t, p)$ as a function of time, $t$, described by the Boltzmann equation,
\begin{equation}
  \frac{\D f_\chi}{\D t}={\cal C}[f_\chi]
  \label{eq:fcoll}
\end{equation}
where ${\cal C}[f_\chi]$ is the collision operator. For freeze-in and superWIMP production -- both arising from decays of the form $B\to A \chi$ -- the collision operator involves the very same interaction, but with different assumptions regarding the momentum distribution of the mother particle $B$. For freeze-in, $B$ is assumed to be in thermal equilibrium with the thermal bath, while superWIMP production denotes the late decay of $B$ after it has chemically decoupled. 

While the resulting relic density depends on the details of the $B$ freeze-out, we can find approximate analytic expressions for the $n$-th moments of the momentum distribution that characterize their impact on structure formation.
For freeze-in from decays and superWIMP production, we find
\begin{equation}
  \qnav|_{\rm FI,\,dec}\simeq \frac{4}{3\sqrt{\pi}}\Gamma\left(\frac{5}{2}+n\right)\times \delta^{n}\quad \text{and}\quad
  \label{eq:qavdec}
\qnav|_{\rm SW}\simeq  \q 2 R^{\rm SW}_\Gamma\w^{-n/2}\delta^n\,\Gamma\left(\frac{n}{2}+1\right)\, ,
\end{equation}
respectively. In eq.~\eqref{eq:qavdec}, $\Gamma$ is the mathematical Gamma function, $q=p/T$ is the momentum mode (where $p$ is the absolute value of the spatial momentum and $T$ the temperature), 
$\delta$ parametrizes the relative mass splitting, $\delta=(m_B^2-m_A^2)/m_B^2$, and $R_\Gamma^{\rm SW}$ is the decay rate, $\Gamma_{B\to A\chi}$, conveniently rescaled:
\begin{equation}
R_\Gamma^{\rm SW}=\frac{M_{\rm Pl} \Gamma_{B\to A\chi}}{m_B^2} \sqrt{\frac{45}{4 \pi^3 g_*(\TSW)}}\,,
  \label{eq:Rgam}
\end{equation}
where $M_{\rm Pl}$ is the Planck mass, and $g_*(\TSW)$ denotes the number of relativistic degrees of freedom in the thermal bath contributing to the radiation energy density at the time of superWIMP decay.

As the scales considered by Lyman-$\alpha$ data lie in the non-linear regime of structure formation, a derivation of DM constraints usually requires computationally expensive hydrodynamic simulations.
However, to good approximation, we can use the results obtained for warm DM and re-interpret them considering the linear matter power spectrum which we obtain from a modified version of \textsc{class}~\cite{Blas:2011rf, Lesgourgues:2011rh}. 
In~\cite{Decant:2021mhj}, the analysis has been performed following three different strategies using the velocity dispersion as done in~\cite{Bae:2017dpt}, an analytical fit to the transfer function following~\cite{Bode:2000gq,Viel:2005qj}, and using the area criterion~\cite{Murgia:2017lwo,Schneider:2016uqi}. The three methods have led to compatible results. Here we report the results we obtain using the latter method. 
To this end, we compute the integral over the one-dimensional linear matter power spectrum (the area $A$)
for the warm DM benchmark model corresponding to the $95\,\%$ C.L. limit, $m_{\rm WDM}^{\text{Ly}-\alpha}=5.3\,$keV. 
Performing the same computation for the test model and comparing the respective results leads to the constraints
\begin{equation}
m_\chi \gtrsim 
   \begin{dcases*}
 15 \, {\rm keV} \times \delta  \, \left( \frac{\gs(\TFI)}{106.75} \right)^{1/3} & for freeze-in through decays,\\
3.8 \, {\rm keV}\times \delta \, \left( \frac{\gs(\TSW)}{106.75} \right)^{1/3} \left(R_\Gamma^{\rm SW}\right)^{-1/2}& for superWIMP\,.\\
   \end{dcases*}
   \label{eq:arealim}
\end{equation}
Here, $\gs(\TSW)$ denote the relativistic degrees of freedom contributing to the entropy density.

An advantage of the area criterion is its applicability to the mixed scenario where similar contributions stem from the freeze-in and superWIMP production. We consider this case in the next section.

\section{Application to a simplified model}\label{sec:appl}

For an application of the analysis described above, we
consider a simplified $t$-channel mediator DM model. It supplements the standard model with a Majorana fermion, $\chi$, and a colored scalar mediator, $\tilde t$, with gauge quantum numbers identical to the right-handed top quark. An imposed $Z_2$ symmetry stabilizes the DM candidate $\chi$ for $m_\chi<m_{\tilde t}$. 
The interactions are described by the Lagrangian
\begin{equation}
    \mathcal{L}_\text{int} = |D_\mu \tilde t|^2 + \lambda_\chi \tilde t\, \bar{t}\,\frac{1-\gamma_5}{2}\chi +\text{h.c.} 
    \label{eq:stopmodel}
\end{equation}
where $D_\mu$ is the covariant derivative and $t$ the top quark Dirac field. 
The masses $m_\chi,m_{\tilde t}$ and the coupling $\lambda_\chi$ are considered to be free model parameters. 

\begin{figure*}[t]
  \begin{center}
    \includegraphics[width=0.4185\textwidth,trim= {0.cm 0.cm 0.cm 0.13cm},clip]{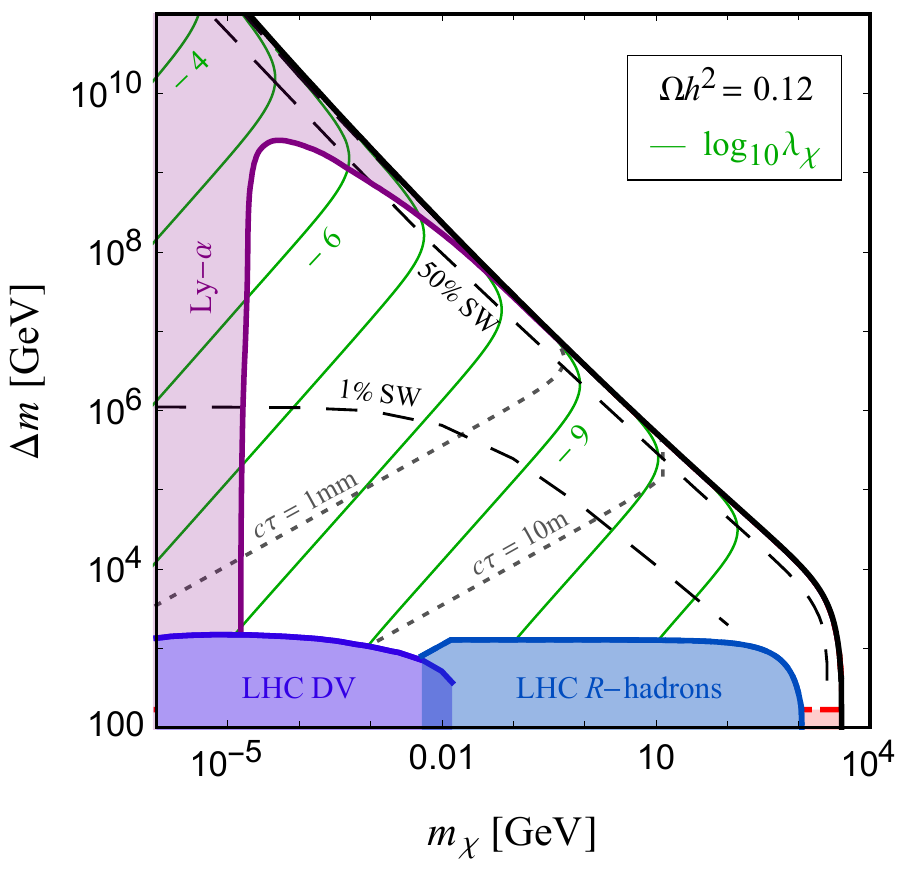}
    \hspace*{6mm}
    \includegraphics[width=0.41013\textwidth,trim= {0.cm 0.cm 0.cm 0.13cm},clip]{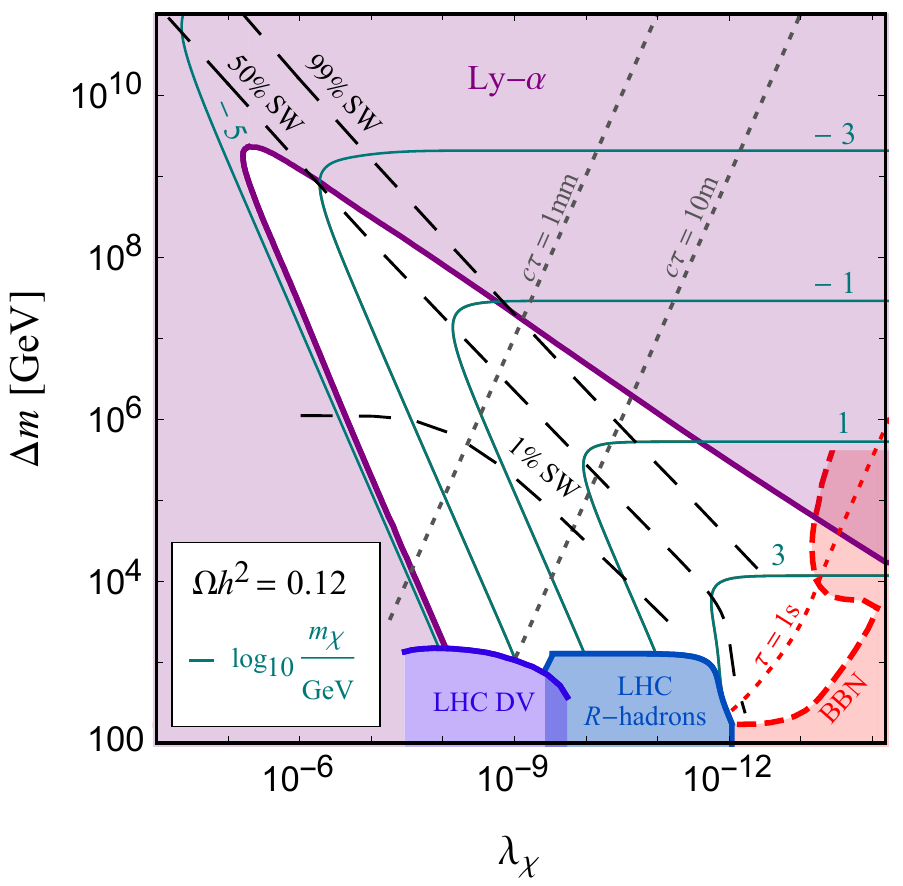}
  \vspace*{-5mm}
  \end{center}
\caption{Cosmologically viable parameter space of the considered $t$-channel mediator model. \textbf{Left}:~Projection onto the plane spanned by $m_\chi$ and $\Delta m= m_{\tilde t} - m_\chi$. The green contours denote decades of the coupling $\lambda_\chi$. For parameter points to the right of the black, thick line, DM is overabundant regardless of the coupling, \emph{i.e.}~no solution can be found. \textbf{Right}:~Projection onto the $m_\chi$-$\lambda_\chi$-plane. The cyan contours denote decades of $m_\chi/$GeV. (To reduce clutter we only display every second line.) Note that the scale of the abscissa has been reversed allowing for a better comparison. In both panels, the black, long-dashed curves denote contours of equal superWIMP (SW) contribution to the total relic density. The gray dotted lines denote contours of equal decay length. Our constraints from the Lyman-$\alpha$ observations (Ly-$\alpha$) are shown in purple, while BBN bounds~\cite{Jedamzik:2006xz} are displayed in red. Constraints from LHC searches for displaced vertices (DV)~\cite{Calibbi:2021fld,ATLAS:2017tny} and $R$-hadrons~\cite{ATLAS:2019gqq} are shown in royal blue and aqua blue, respectively, see~\cite{Decant:2021mhj} for further details.
}
\label{fig:paramspace}
\end{figure*}
\begin{figure}[b]
  \begin{center}
    \includegraphics[width=.46\textwidth,trim= {0.cm 0.07cm 0.cm 0.22cm},clip]{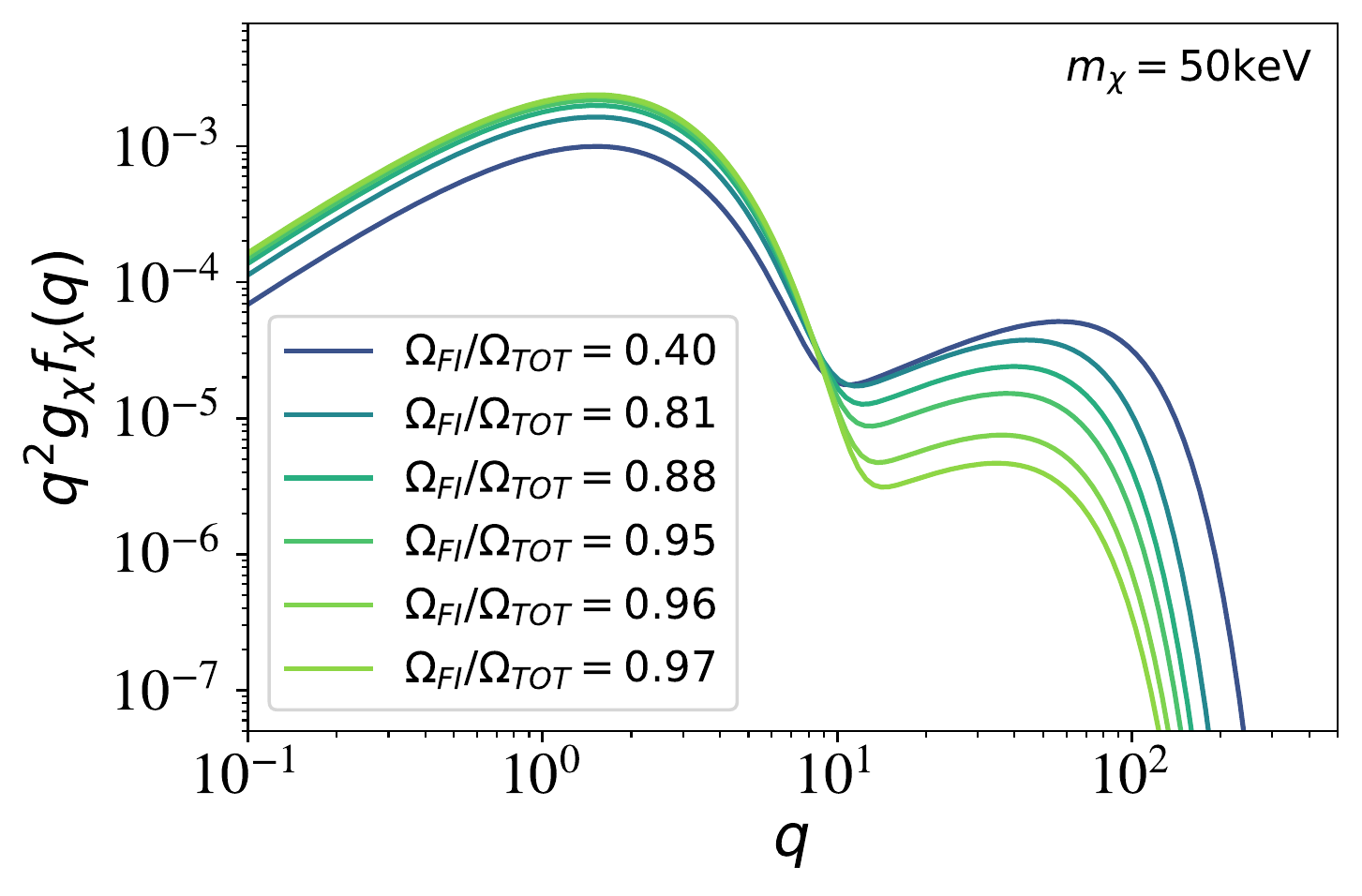}
     \hspace*{6mm}
    \includegraphics[width=.434\textwidth,trim= {0.cm 0.07cm 0.cm 0.22cm},clip]{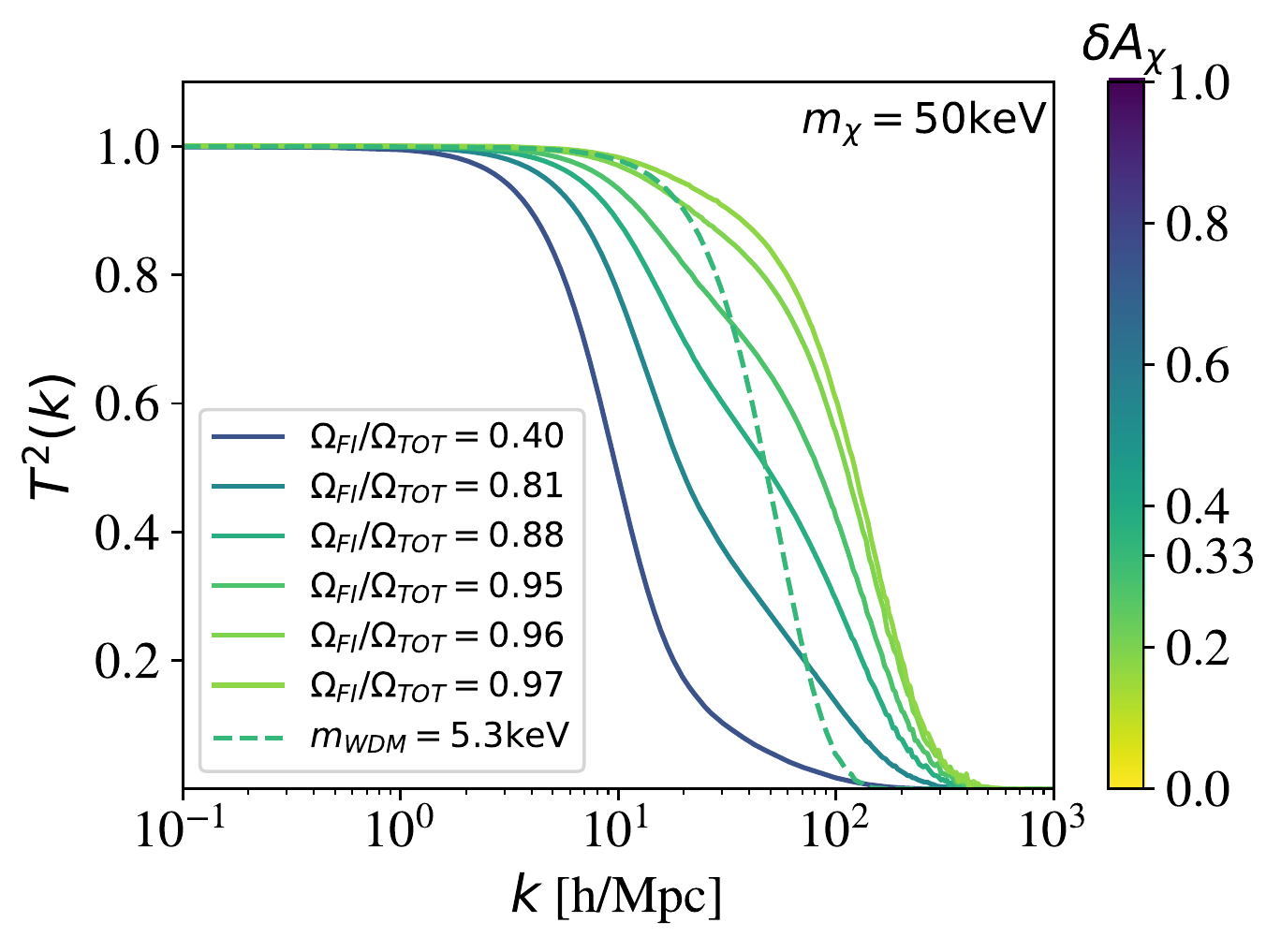}
  \vspace*{-5mm}
  \end{center}
\caption{Momentum distribution (left panel) in the mixed freeze-in/superWIMP scenario for different relative freeze-in contributions and their respective transfer functions (right panel). The color code represents the relative difference of the respective area $A_\chi$ to the one for cold DM, $\delta A_\chi = 1- A_\chi/A_{\text{CDM}}$.}
\label{fig:transfers}
\end{figure}

We assume that the model explains the measured relic density, $\Omega_\chi h^2|_\text{FI}(\lambda_\chi) + \Omega_\chi h^2|_\text{SW}(\lambda_\chi)=0.12$, allowing us to compute the required $\lambda_\chi$ for every mass point, see~\cite{Decant:2021mhj} for further details. The allowed parameter space is shown in Fig.~\ref{fig:paramspace}. While LHC searches for long-lived particles and bounds from Big Bang Nucleosynthesis (BBN) exclude the parameter space towards small mediator masses, Lyman-$\alpha$ constraints corner the parameter space towards small DM masses, small couplings and large mediator masses. In the latter region, the freeze-in and superWIMP contribution are of similar size (see long-dashed contours). The respective momentum distribution and transfer functions are shown in Fig.~\ref{fig:transfers}. Considering $m_\chi=50\,$keV, we choose mediator masses around the Lyman-$\alpha$ exclusion limit. They are characterized by different relative contributions from freeze-in and superWIMP production. These contributions give rise to the first and second bump in the momentum distribution, respectively. The transfer function for sizable admixtures show considerable deviations in shape from the warm DM case shown as the dashed curve.

\vspace{-0.1ex}

\section{Conclusion}\label{sec:concl}

\vspace{-0.3ex}

Lyman-$\alpha$ forest observations are an interesting probe of the DM momentum distribution and, hence, may provide insights into the underlying DM genesis mechanism. We reinterpreted warm DM bounds for non-thermalized DM, \ie~DM produced via the freeze-in and superWIMP mechanism or an admixture of both. To do so, we employed the area criterion considering the integral over the one-dimensional linear matter power spectra as a measure. For the computation of linear matter power spectra we employed a modified version of \class\@.
We obtained approximate analytic expressions for the pure scenarios while demonstrating the mixed freeze-in/superWIMP scenario in a numerical example. To this end, we presented an application to a top-philic $t$-channel mediator model. Its cosmologically viable parameter space is cornered from all sides by LHC and BBN bounds (towards small mediator masses) and by Lyman-$\alpha$ constraints (towards small DM masses, small DM couplings and large mediator masses). In the region of similar contributions form both production mechanisms, the linear power spectrum significantly deviates in shape from the one of the canonical warm DM scenario 
motivating further studies beyond the area criterion, such as a dedicated analysis using a full Lyman-$\alpha$ likelihood.

\providecommand{\href}[2]{#2}\begingroup\raggedright\endgroup

\end{document}